\documentclass[10pt,final, conference]{IEEEtran}
\usepackage{amsmath,epsfig, pstricks}

\IEEEoverridecommandlockouts

\newcommand{\SINR}{\mbox{SINR}}

\newcommand{\Ind}[1]{1_{\left\{ #1 \right\}}}

\newtheorem{theorem}{Theorem}

\def\blfootnote{\xdef\@thefnmark{}\@footnotetext} 

\IEEEtriggeratref{2}
\begin{document}			   
\title{On the Spectral Efficiency of Links with Multi-antenna Receivers in Non-homogenous Wireless Networks}                      

\author{\IEEEauthorblockN{Siddhartan Govindasamy}
\IEEEauthorblockA{F. W. Olin College of Engineering\\
Needham, MA, USA\\
Email: siddhartan.govindasamy@olin.edu}
\and
\IEEEauthorblockN{Daniel W. Bliss
$^\dagger$}
\IEEEauthorblockA{MIT Lincoln Laboratory\\
Lexington, MA,  USA\\
Email: bliss@ll.mit.edu}\thanks{\noindent $\dagger$ Daniel W. Bliss is currently working at MIT Lincoln Laboratory. Opinions, interpretations, conclusions, and recommendations are those of the authors and are not necessarily endorsed by the United States Government.}
}

\maketitle 

\begin{abstract}
An asymptotic technique is developed to find the Signal-to-Interference-plus-Noise-Ratio (SINR) and spectral efficiency of a link with N receiver antennas in wireless networks with  non-homogeneous distributions of nodes. It is found that with appropriate normalization, the SINR and spectral efficiency converge with probability 1 to asymptotic limits as N increases. This technique is applied to networks with power-law node intensities, which includes homogeneous networks as a special case, to find a simple approximation for the spectral efficiency.  It is found that for receivers in dense clusters, the SINR grows with N at rates higher than that of homogeneous networks and that constant spectral efficiencies can be maintained if the ratio of N to node density is constant. This result also enables the analysis of a new scaling regime where the distribution of nodes in the network flattens rather than increases uniformly. It is found that in many cases in this regime, N needs to grow approximately exponentially to maintain a constant spectral efficiency. In addition to strengthening previously known results for homogeneous networks, these results provide insight into the benefit of using antenna arrays in non-homogeneous wireless networks, for which few results are available in the literature.
\end{abstract}

\section{Introduction}
Multiple antenna systems can improve the performance of point-to-point wireless links by increasing data rates  through spatial multiplexing and array gain as well improving robustness through diversity. In wireless networks, multi-antenna arrays can be used to mitigate interference enabling simultaneous transmissions. The data rates achievable using antenna arrays for interference mitigation is highly dependent on the distribution of nodes in space which impacts inter-node distances, signal and interference levels, and hence data rates.

 Thus, the spectral efficiencies achievable in wireless networks with spatially-distributed multi-antenna nodes have received much attention.  Chen and Gans \cite{ChenGans} found that in order to maintain constant per-link data rates in ad-hoc wireless networks, it is necessary to increase the number of receiver antennas linearly with the number of users. Govindasamy et. al. \cite{JSACPaper} found that the SINR of a representative link in a wireless network where signal power decays according to a power-law with path-loss exponent $\alpha$, grows as the ratio of the number of antennas $N$ to the user density $\rho$ raised to the power $\alpha/2$. Jindal et. al. \cite{JindalRethinking} found that it is possible to increase the area spectral efficiency of wireless networks linearly with the number of receiver antennas by increasing the density of simultaneous transmissions using a partial zero-forcing receiver. Hunter et. al. \cite{HunterAndrews}  and Kountouris and Andrews \cite{Kountouris} consider several transmit encoding approaches with the former considering approaches such as antenna selection, maximal ratio transmissions and space-time orthogonal block codes,  and the latter analyzing dirty-paper coding, zero-forcing and block-diagonalization. Ali et. al.  \cite{GagnonMMSE}, and Louie et. al. \cite{JindalMultiStream} found the CDF of the SINR in networks with Linear Minimum-Mean-Square-Error (MMSE) receivers in Rayleigh fading with single and multi-stream transmissions respectively. Vaze and Heath \cite{VazeHeath}, Hunter and Andrews \cite{HunterAndrewsMultStream} and Govindasamy et. al. \cite{AsilomarTxCSI} considered multi-stream transmission with channel-state-information (CSI) at transmitters using  different assumptions on the transmitter and receiver processing. These works assume a uniformly random distribution of nodes on the plane where the intensity of the nodes does not vary with location. Networks where node intensities vary with location  arise in many practical situations such as when users congregate at hot-spots. For non-homogeneous distribution of nodes, Ganti and Haenggi \cite{GantiClustered} develop techniques to analyze wireless networks with clusters of users and Tresch and Guillaud  in \cite{TreschCluster} analyze interference alignment in such networks. Win et. al. \cite{WinNetworkInterference} develop a general set of tools to analyze interference in wireless networks with possibly  non-homogeneous node distributions.

\begin{figure}
\centering
\includegraphics[width = 2.5in]{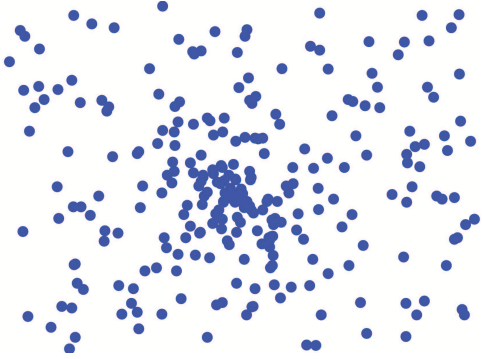}
\caption{Randomly distributed points with linear radial intensity}\label{Fig:LinIntens}
\end{figure}

In this work we develop an asymptotic technique to find the spectral efficiency and SINR of a link with multiple antennas at the receiver, in a network with a non-homogeneous distribution of nodes in the interference-limited regime, with Linear MMSE receivers.  In contrast to \cite{GantiClustered}  and \cite{TreschCluster}, we assume that the intensity function is fixed which can arise in situations where the locations of hot-spots are known. The spatial distribution of nodes plays an important role in the rate at which SINR grows with the number of antennas at the receiver and hence strongly influences the trade-off between the increased hardware costs of antenna arrays and increased data rates. For instance, we find that in a network where the intensity of users decays inversely with the distance from the receiver (as illustrated in Figure \ref{Fig:LinIntens}), the SINR grows as $N^\alpha$ compared to $N^{\alpha/2}$ for a uniform intensity of nodes. This result indicates that antenna arrays can provide an even higher increase in SINR in such networks compared to uniformly random networks.

We address the added complexity of non-uniform node intensities by using an asymptotic analysis, taking the radius of a circular network and number of receiver antennas to infinity. Monte-Carlo simulations used to support our findings indicate that the results are useful for systems with moderate numbers of antennas which is consistent with other works on multi-antenna systems utilizing infinite random matrix theory \cite{DanBlissEnv}. Furthermore, as hardware costs decrease and transmissions move to higher frequencies, it will be possible to fit large numbers of antennas on standard mobile devices. For instance, at a nominal carrier frequency of 6GHz, it is possible to fit 30 or more antenna elements on a standard laptop computer with wavelength separation ($\sim 5$ cm at 6GHz). 

We start by showing that with appropriate normalization, the SINR of a link located at the center of a network converges with probability 1 to an asymptotic limit that is dependent on an intensity function $\Lambda(r, \theta)$ where $r \Lambda(r, \theta)dr d\theta$ approximately equals the average number of points that occur in a small rectangle of sides $dr$ and $r d\theta$ around the point $(r,\theta)$. We apply this result to power-law intensity functions where $\Lambda(r, \theta) = \rho r^{\epsilon}$ for some $\epsilon$, to find an approximation for the SINR when $N$ is large. This result which includes the uniform density as a special case (i.e. $\epsilon = 0$) can be used to study the scaling behavior of the SINR as the user density increases. In addition to the case where the density of users increases uniformly ($\rho \to \infty$), this result also enables us to study a scaling regime  where the user density ``flattens'' i.e. $\epsilon \to 0$ which can be used to model networks where the user distribution approaches a homogeneous intensity. 

 Additionally, since we make no assumptions on the distribution of the channels between pairs of antennas, our results can be adapted to  Direct-Sequence Code-Division-Multiple-Access (DS/CDMA) systems with random spreading codes.

We present the system model in Section II followed by the main results for general intensity functions and power law intensity functions in Section III. Section IV contains numerical results followed by a summary and conclusions. Outlines fof the derivations of the main results are presented in the appendix. 
 
\section{System Model}

Consider a circular network of radius $R$ centered at the origin. Assume that $n$ transmitting  nodes are distributed independently and randomly in the circle and that the $i$-th node is at location $(r_i,\theta_i)$ where the angle and distance are distributed jointly according to
\begin{align}
f_{r,\theta} (r, \theta) = r \Lambda(r, \theta).
\end{align}
Additionally, assume that $n$ and $R$ are related by the following equation:
\begin{align}
n = \int_0^{R}\int_{0}^{2\pi}r\Lambda(r,\theta)\, dr d\theta. \label{Eqn:NodesRadiusRel}
\end{align}
which we assume is invertible with $g(n) = R$ and that $\Lambda(r, \theta)$ is such that $n\to \infty$ as $R \to\infty$, i.e. there are an infinite number of nodes in an infinite radius network. 

We shall refer to these $n$ transmitting nodes as interferers. Assume that a receiver is located at the origin and is communicating with an \emph{additional} transmitter located at a fixed distance $r_T$ away. We shall refer to this transmitter as the target transmitter. The interferers are assumed to be communicating with other receivers whose locations do not impact the results. 

Suppose that all nodes transmit with unit power and that the average  power received at the origin from node-$i$, $p_i$ follows the standard inverse power-law model
\begin{align}
p_i = r_i^{-\alpha}
\end{align}
with $\alpha > 2$. 
Each transmitting node has a single antenna and the receiver at the origin has $N$ antennas. If the $N\times 1$ vector of received samples at each antenna of the receiver at a given time is contained in the vector $\mathbf{y}$, the system can be described by the following linear equation
\begin{align}
\mathbf{y} = \bar{\mathbf{H}}\bar{\mathbf{P}}^{\frac12} \mathbf{x} + \mathbf{w} \label{Eqn:LinearSystem}
\end{align}
where the $N\times (n+1)$ matrix $\bar{\mathbf{H}}$ contains independent, identically-distributed (IID) complex random variables of unit variance. Note that we do not require the channel co-efficients to be Gaussian. $\bar{\mathbf{P}} = diag(r_T^{-\alpha}, r_1^{-\alpha}, \cdots r_n^{-\alpha})$ and $\mathbf{x}$ is an $(n+1)\times 1$ vector of transmit samples from the target transmitter and $n$ interferers. $\mathbf{w}$ is a vector of IID noise samples with variance $\sigma^2_N=\sigma^2N g(n)^{-\alpha}$. We define the noise in this manner to enable an asymptotic analysis in the interference-limited regime as the number of receiver antennas $N \to \infty$. Without this normalization, as $N\to\infty$, the MMSE receiver will be able to suppress interference to levels comparable to the thermal noise resulting in the system no longer being interference-limited. Since our focus in this work is on the interference-limited regime, the actual value of the thermal noise power will not be relevant in the results.  Additionally, we define a  normalized version of the SINR $\beta_N$ which we show converges with probability 1 in the next section
\begin{align}
\beta_N = r_T^{\alpha}(g(Nc))^{-\alpha}\SINR
\end{align}
where $c = n/N$.

The receiver uses a linear-MMSE estimator to estimate the sample from the target transmitter, i.e. the first entry of $\mathbf{x}$. The SINR associated with this estimator is known to equal
\begin{align}
\mbox{SINR} = \mathbf{h}_T^\dagger\left(\mathbf{HPH}^\dagger + \sigma_N^2\mathbf{I}\right)^{-1}\mathbf{h}_T.
\end{align}
where $\mathbf{h}_T$ is the first column of $\bar{\mathbf{H}}$, $\mathbf{P} = diag(p_1, p_2, \cdots p_n)$ and $\mathbf{H}$ is equal to $\bar{\mathbf{H}}$ with the first column removed. 

\section{Main Results}

\subsection{General Intensity Functions}
\begin{theorem}\label{Theorem:MainTheorem}
Let
\begin{align}
 H(x) = 1-x^{-\frac{2}{\alpha}}\lim_{N\to\infty}\frac{\int_{0}^{2\pi}\Lambda(g(Nc) x^{-\frac{1}{\alpha}},\theta)\Ind{1<x < \infty}\, d\theta}{\int_{0}^{2\pi}\Lambda(g(Nc),\theta)\,d\theta}.\nonumber
\end{align} 
Then, in the limit  as $n, N, R\to \infty$ such that $c = n/N$ and \eqref{Eqn:NodesRadiusRel} always holds, $\beta_N \to \beta$ with probability 1, where  $\beta$ is a unique non-negative solution to the following equation:
\begin{align}
-\beta \sigma^2 + 1 = \beta c\int_0^\infty \frac{x \, dH(x)}{1+x\beta} \label{Eqn:FixedPoint}
\end{align}
\end{theorem}
Thus, when the number of antennas at the  receiver is large, the SINR can be approximated by removing the scaling as follows:
\begin{align}
\SINR \approx r_T^{-\alpha}\beta g(n)^{\alpha} =  r_T^{-\alpha}\beta g(Nc)^{\alpha}.
\end{align}
Note that in many cases of interest (e.g. the power law intensity function described below), the RHS above is does not depend strongly  $c$.

\begin{theorem} \label{Theorem:SpecEff}
If $(g(n))^\alpha$ is a function of lower than exponential order, i.e. for all $A > 1$, $A^{-n}(g(n))^\alpha \to 0$ as $n \to \infty$, the following hold with probability 1:
\begin{align}
\log_2(1+\SINR)-\log_2(1+r_T^{-\alpha}(g(Nc))^\alpha\beta) \to 0
\end{align}
and
\begin{align}
E[\log_2(1+\SINR)] - \log_2(1+r_T^{-\alpha}(g(n))^\alpha\beta)  \to 0.
\end{align}
\end{theorem}
The latter two results are useful to characterize the spectral efficiency when Gaussian code-books are used.

\subsection{Power-Law Intensity}
Now consider an intensity function $\Lambda(r, \theta) = \rho r^\epsilon$ for $\epsilon > -2$.  This model can be used to approximate systems where the intensity of nodes decays with distance from the origin such as that depicted in Figure \ref{Fig:LinIntens}.

\begin{theorem} \label{Theorem:PowerLaw}
Under the conditions of Theorem \ref{Theorem:MainTheorem}, $\beta_N \to \beta$ with probability 1, where $\beta$ is a unique positive solution to:

\begin{align}
&-\beta\sigma^2 + 1 = \frac{(2+\epsilon) c}{\alpha}\beta^{\frac{2+\epsilon}{\alpha}}\pi\csc\left(\frac{2\pi+\pi\epsilon}{\alpha}\right)\nonumber \\
& -\frac{(2+\epsilon)\beta c}{\alpha-2-\epsilon}\,_2F_1\left(1, \frac{-2-\epsilon+\alpha}{\alpha}, \frac{-2-\epsilon+2\alpha}{\alpha}, -\beta \right)
\end{align}
\end{theorem}

With $\epsilon  = 0$, which corresponds to a uniform intensity of nodes, the result above is a strengthening of an earlier work \cite{JSACPaper} where it was shown that $E[\beta_N] \to \beta$ and var$\left\{\beta_N\right\} \to 0$.  

For $\alpha - 2-\epsilon > 0$, we can find a simple approximation for $\beta$ in the expression above if we write $\eta_N = \beta_Nc^{\frac{\alpha}{2+\epsilon}} = \left(\frac{2+\epsilon}{2\pi \rho}N\right)^{-\frac{\alpha}{2+\epsilon}}r_T^\alpha\, \SINR$ and $\eta = c^{\frac{\alpha}{2+\epsilon}}\beta$, we have:
\begin{align}
&-\eta c^{-\frac{\alpha}{2+\epsilon}}\sigma^2 + 1  = \frac{(2+\epsilon)}{\alpha}\eta^{\frac{2+\epsilon}{\alpha}}\pi\csc\left(\frac{2\pi+\pi\epsilon}{\alpha}\right)\nonumber \\
& \;\;\;\;\;\,-\frac{(2+\epsilon)\eta c^{\frac{2+\epsilon-\alpha}{2+\epsilon}}}{\alpha-2-\epsilon} \times \nonumber \\
&\;\;\;\;\;\;\;\,_2F_1\left(1, \frac{-2-\epsilon+\alpha}{\alpha}, \frac{-2-\epsilon+2\alpha}{\alpha}, -\eta c^{-\frac{\alpha}{2+\epsilon}} \right) \nonumber
\end{align}

Using Euler's Hypergeometric transforms yields:
\begin{align}
&-\eta c^{-\frac{\alpha}{2+\epsilon}}\sigma^2 + 1  = \frac{(2+\epsilon)}{\alpha}\eta^{\frac{2+\epsilon}{\alpha}}\pi\csc\left(\frac{2\pi+\pi \epsilon}{\alpha}\right)\nonumber \\
& -\frac{(2+\epsilon)}{\alpha-2-\epsilon}\frac{\eta c^{\frac{2+\epsilon-\alpha}{2+\epsilon}}}{1+\eta c^{\frac{-\alpha}{2+\epsilon}}}\,_2F_1\left(1,1, \frac{-2-\epsilon+2\alpha}{\alpha}, \frac{\eta c^{\frac{-\alpha}{2+\epsilon}}}{1+\eta c^{\frac{-\alpha}{2+\epsilon}}} \right)\nonumber
\end{align}
Since $\alpha - 2-\epsilon > 0$, the parameters of the hypergeometric function above are positive implying that it is a monotonically increasing function of the argument. Hence,
 \begin{align*}
&  \left | \frac{1} {(1+\eta c^{\frac{-\alpha}{2+\epsilon}})}\,_2F_1\left(1,1, \frac{-2-\epsilon+2\alpha}{\alpha}, \frac{\eta c^{\frac{-\alpha}{2+\epsilon}}}{1+\eta c^{\frac{-\alpha}{2+\epsilon}}} \right) \right | \leq \\
 &\,_2F_1\left(1,1; \frac{-2-\epsilon+2\alpha}{\alpha},1 \right) = \frac{2+\epsilon-\alpha}{2+\epsilon}
\end{align*}
For large $c$, i.e. when the ratio of nodes in the network to the number of antennas at the receiver is high and small $\sigma^2$:
\begin{align}
& \frac{(2+\epsilon)}{\alpha}\eta^{\frac{2+\epsilon}{\alpha}}\pi\csc\left(\frac{2\pi+\pi\epsilon}{\alpha}\right)\approx 1
\end{align}
Thus for large $N$, the SINR can be approximated by
\begin{align}
\SINR \approx r_T^{-\alpha}\left(\frac{\alpha}{2\pi^2}\sin\left(\frac{2\pi+\pi\epsilon}{\alpha}\right)\right)^{\frac{\alpha}{2+\epsilon}}\left(\frac{N}{\rho}\right)^{\frac{\alpha}{2+\epsilon}} \label{Eqn:SINRApproxPowerLaw}
\end{align}
and by Theorem \ref{Theorem:SpecEff} the spectral efficiency assuming Gaussian codebooks can be approximated as:
\begin{align}
\gamma \approx \log_2\left(1+r_T^{-\alpha}\left(\frac{\alpha}{2\pi^2}\sin\left(\frac{2\pi+\pi\epsilon}{\alpha}\right)\right)^{\frac{\alpha}{2+\epsilon}}\left(\frac{N}{\rho}\right)^{\frac{\alpha}{2+\epsilon}}\right). \label{Eqn:SpecEffApproxPowerLaw}
\end{align}
Re-writing the above equation in terms of $N$ yields
\begin{align}
N\approx \rho r_T^{2+\epsilon}\left(2^\gamma -1\right)^{\frac{2+\epsilon}{\alpha}}\left(\frac{2\pi^2}{\alpha}\csc\left(\frac{2\pi+\pi\epsilon}{\alpha}\right)\right).\label{Eqn:NFunce}
\end{align}
This indicates that to maintain a constant spectral efficiency $N$ must increase linearly with $\rho$ and $r^{2+\epsilon}$ but approximately \emph{exponentially} with $\epsilon$.

\section{Numerical Results}

We performed Monte-carlo simulations of network topologies to validate the results and approximations from the previous sections. The receiver was placed at the origin and 10000 nodes were placed at random polar coordinates in a disk of radius $R$ which satisfies \eqref{Eqn:NodesRadiusRel} with $\rho = 0.01$. The distances were drawn from a density proportional to $r^{\epsilon+1}$ between $0$ and $R$ and the angles were uniform in $(0,2\pi]$. The channels between the antennas of a node at a given distance and the  receiver were IID, circularly symmetric Gaussian random variables. We used unit transmit power and a \emph{constant}  noise power of $10^{-14}$ and the target transmitter was at distance $r_T = 10$. Note that in this regime, the system is interference limited and the specific value of the noise power does not appreciably impact the spectral efficiency. 1000 trials of each set of parameters was simulated and the spectral efficiencies were computed using the Shannon formula. We considered $\alpha = 2.5,3.5, 3, 4,$ and $4.5$ and $\epsilon = -1, -0.75, -0.5, -0.25$, and $0$.

\begin{figure}
\begin{center}
\includegraphics[width = 3.5in]{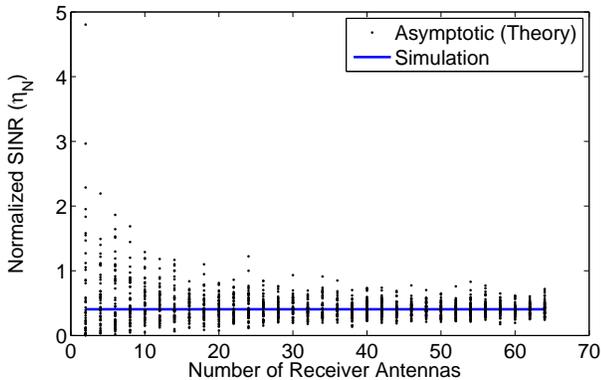}
\caption{Normalized SINR vs. number of receiver antennas for power-law intensity with $\epsilon = -0.5$ and $\alpha = 3$. The points are 50 random samples out of the simulated values of  $\eta_N = \left(\frac{2+\epsilon}{2\pi \rho}N\right)^{-\frac{\alpha}{2+\epsilon}}r_T^\alpha\, \SINR$ , and the solid lines indicate the asymptotic approximation. }\label{Fig:NormSINRPowerLaw0_5}
\end{center}
\end{figure}

The points in Figure \ref{Fig:NormSINRPowerLaw0_5} illustrate a random sampling of 50 points from the simulation of $\eta_N = \left(\frac{2+\epsilon}{2\pi \rho}N\right)^{-\frac{\alpha}{2+\epsilon}}r_T^\alpha\, \SINR$ for each value of $N$, and the solid line shows the asymptotic prediction for a system with $\epsilon = -0.5$ and $\alpha = 3$. We plotted a small subset of the simulated points for clarity. The concentration of points with increasing $N$ indicates convergence of the normalized SINR to the asymptotic value.

\begin{figure}
\begin{center}
\includegraphics[width = 3.5in]{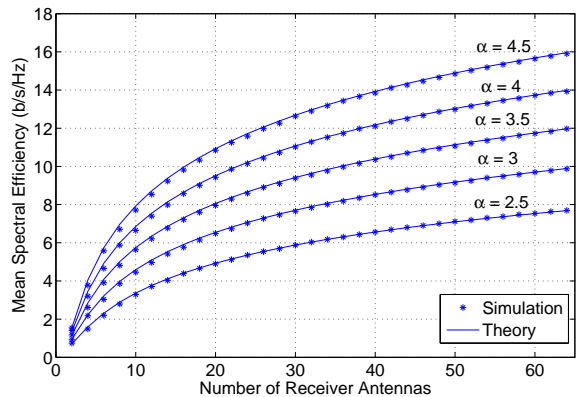}
\caption{Mean spectral efficiency vs. number of receiver antennas for power-law intensity with $\epsilon = -0.5$ and varying $\alpha$. The markers indicate simulated values and the solid lines indicate the theoretical approximation of \eqref{Eqn:SpecEffApproxPowerLaw}. }\label{Fig:MeanSpecEffPowerLaw0_5}
\end{center}
\end{figure}

\begin{figure}
\begin{center}
\includegraphics[width = 3.5in]{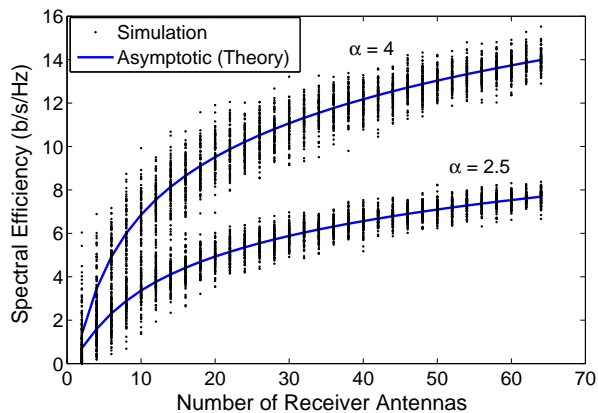}
\caption{Simulated spectral efficiencies vs. number of receiver antennas for power-law intensity with $\epsilon = -0.5$ and different values of $\alpha$. The points indicate a random sampling of 50 simulated values, and the solid lines indicate the theoretical approximation. }\label{Fig:SpecEffPowerLaw0_5Scatter}
\end{center}
\end{figure}

Figures \ref{Fig:MeanSpecEffPowerLaw0_5} and \ref{Fig:SpecEffPowerLaw0_5Scatter} show the spectral efficiencies for $\epsilon  = -0.5$ and different values of $\alpha$. Figure \ref{Fig:MeanSpecEffPowerLaw0_5} shows the simulated mean spectral efficiencies (markers) and the asymptotic prediction of \eqref{Eqn:SpecEffApproxPowerLaw}. In all the cases shown, the mean spectral efficiency is within $5\%$ of the asymptotic value for $N > 6$ which illustrates the utility of \eqref{Eqn:SpecEffApproxPowerLaw} for reasonably small numbers of receiver antennas. Figure \ref{Fig:SpecEffPowerLaw0_5Scatter} shows a random sampling of 50 trials of the spectral efficiency for each value of $N$, for $\alpha = 2.5$ and $\alpha = 4$. The concentration of points indicates convergence of the spectral efficiency to its asymptotic value.

\begin{figure}
\begin{center}
\includegraphics[width = 3.5in]{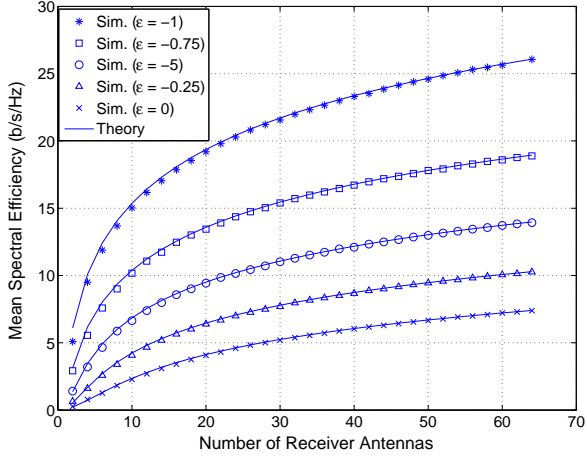}
\caption{Mean spectral efficiency vs. number of antennas for different values of $\epsilon$ with $r_T = 10, \alpha = 4, B = 0.01$. }\label{Fig:SpecEffMulte}
\end{center}
\end{figure}

\begin{figure}
\begin{center}
\includegraphics[width = 3.5in]{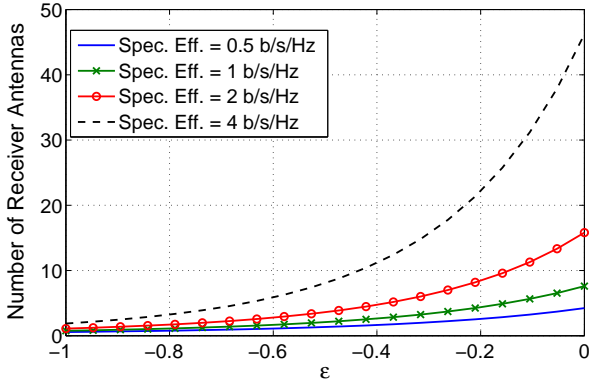}
\caption{Number of receiver antennas required to maintain constant spectral efficiencies vs. $\epsilon$. Link-length $r_T = 10, \alpha = 3, B = 0.01$ }\label{Fig:Flattening}
\end{center}
\end{figure}

Figure \ref{Fig:SpecEffMulte} shows the mean spectral efficiency for $\alpha = 4$ and different values of $\epsilon$. The mean spectral efficiency is within $5\%$ of the asymptotic value for $N > 6$ indicating that \eqref{Eqn:SpecEffApproxPowerLaw} holds for a range of different $\epsilon$ and reasonable numbers of antennas. Figure \ref{Fig:Flattening} is a plot of $N$ vs. $\epsilon$ from \eqref{Eqn:NFunce} for different values of the spectral efficiency. Note that an approximately exponential increase in $N$ is required to maintain a constant spectral efficiency with increasing $\epsilon$. Furthermore, large spectral efficiencies greatly impact the rate at which $N$ needs to increase with $\epsilon$. 

\section{Summary and Conclusions}
We have developed asymptotic technique to find the SINR and spectral efficiency  of an interference-limited link with a multi-antenna receiver in a wireless network with nodes distributed according to a non-homogenous point process. We find that with appropriate normalization, the SINR and spectral efficiency of this link converges with probability 1 to an asymptotic limit as the number of receiver antennas and radius of the network grow to infinity.

We applied this result to networks with intensity functions of the form $\rho r^{\epsilon}$ for $\epsilon > -2$, to derive a simple approximation (given by \eqref{Eqn:SpecEffApproxPowerLaw}) for the spectral efficiency. This class of intensity functions can be used to model receivers located in the center of a dense cluster of nodes and includes the homogenous node distribution as a special case ($\epsilon  = 0$). For this model, the SINR is found to be approximately proportional to  $N/\rho$ raised to the power $\frac{\alpha}{2+\epsilon}$ which indicates that per-link spectral efficiency can be held constant by linearly increasing $N$ with density $\rho$, which has been found to hold for homogenous networks in \cite{JSACPaper}. Comparing the $\epsilon = -1$ case for which the SINR grows as $N^\alpha$ and the homogenous case for which the SINR grows as  $N^{\frac{\alpha}{2}}$, we find that the SINR grows with $N$ at a rate significantly faster in the center of a dense cluster compared to a homogenous network. 

Additionally, our results enables the analysis of a new scaling regime where the intensity of the node process flattens ($\epsilon \to 0$) rather than increases uniformly. We find that an approximately exponential growth of $N$ is needed to mantain constant spectral efficiency as $\epsilon \to 0$. 

The asymptotic results are supported by monte-carlo simulations for which the simulated mean spectral efficiencies are close to the asymptotic prediction, differing by less than 5\%  when the number of antennas is greater than 6, for all the cases considered, confirming the the utility of these results for finite numbers of antennas. 

In addition to strengthening previously known results for homogenous networks, these results characterize the  spectral efficiency of a link with multiple receiver antennas  in networks with a non-homogenous distribution of nodes for which there are few results  in the literature. 

\appendix
\subsection{Outline of the derivation for general intensity functions}
We accomplish the normalization of the SINR by $(g(n))^{-\alpha}$ by scaling the interference and  noise powers by $(g(n))^{\alpha}$. Let the scaled interference power due to node $i$ be $\tilde{p}_i =(g(n))^{\alpha}p_i $ and the thermal noise power be $\sigma^2_N (g(n))^{\alpha}$. Consider the CDF of $\tilde{p}_i$ for a given $R$ and recall that  $R = g(n)$.
\begin{align*}
F_{\tilde{p}}(x) &= \mbox{Pr}\left(r_i^{-\alpha} (g(n))^\alpha \leq x \right)= 1-F_r\left(x^{-\frac{1}{\alpha}}g(n)\right) \\
&= 1-\frac{\int_0^{x^{-\frac{1}{\alpha}}g(n)}\int_{0}^{2\pi}r\Lambda(r,\theta) \Ind{r < R}\,dr d\theta}{\int_0^{R}\int_{0}^{2\pi}r\Lambda(r,\theta)\, dr d\theta} \\
&= 1-\frac{\int_0^{R x^{-\frac{1}{\alpha}}}\int_{0}^{2\pi}r\Lambda(r,\theta)\Ind{r < R}\, dr d\theta}{\int_0^{R}\int_{0}^{2\pi}r\Lambda(r,\theta)\, dr d\theta}
\end{align*}
For any positive $x$, as $R\to\infty$, both the numerator and denominator must go to infinity and thus the limit is the ratio of the deriviatives w.r.t. $R$ of the numerator and denominator. Observing that $N\to\infty$ as $R\to\infty$ we have:
\begin{align}
&\lim_{R\to\infty} F_{\tilde{p}}(x) = 1-\frac{\frac{d}{dR}\int_0^{R x^{-\frac{1}{\alpha}}}\int_{0}^{2\pi}r\Lambda(r,\theta)\Ind{r < R}\, dr d\theta}{\frac{d}{dR}\int_0^{R}\int_{0}^{2\pi}r\Lambda(r,\theta)\, dr d\theta}\nonumber \\
 &= 1-x^{-\frac{2}{\alpha}}\lim_{R\to\infty}\frac{\int_{0}^{2\pi}\Lambda(R x^{-\frac{1}{\alpha}},\theta)\Ind{0<x^{-\frac{1}{\alpha}} < 1}\, d\theta}{\int_{0}^{2\pi}\Lambda(R,\theta)\,d\theta} \nonumber \\ 
&= 1-x^{-\frac{2}{\alpha}}\lim_{N\to\infty}\frac{\int_{0}^{2\pi}\Lambda(g(Nc) x^{-\frac{1}{\alpha}},\theta)\Ind{1<x < \infty}\, d\theta}{\int_{0}^{2\pi}\Lambda(g(Nc),\theta)\,d\theta}. \label{Eqn:GeneralLimit}
\end{align}

Let $H(x) = \lim_{R\to\infty} F_p(x)$. By the main result of \cite{BaiSilversteinMIMOCDMA} and the Glivenko-Cantelli theorem, we have $\beta_N \to \beta$ with probability 1, where $\beta$ is a unique non-negative solution to the following equation:
\begin{align}
-\beta \sigma^2 + 1 = \beta c\int_0^\infty \frac{x \, dH(x)}{1+x\beta}
\end{align}
which completes the proof of Theorem \ref{Theorem:MainTheorem}.

To prove the first part of Theorem \ref{Theorem:SpecEff}, consider the regime as $N\to\infty$:
\begin{align*}
& \log_2(1+\SINR)- \log_2(1+r_T^{-\alpha}g(Nc)^\alpha\beta)\\
&= \log_2(1+r_T^{-\alpha}g(Nc)^\alpha\beta_N)- \log_2(1+r_T^{-\alpha}g(Nc)^\alpha\beta)\\
&= \log_2(g(Nc)^{-\alpha}+r_T^{-\alpha}\beta_N)- \log_2(g(Nc)^{-\alpha}+r_T^{-\alpha}\beta)\to 0
\end{align*}
where the last limit holds with probability 1 by the continuity of the log function. The second part can be proved using similar steps as Appendix E of \cite{TxCSIJournal}.

\subsection{Outline of the derivation for power-law intensity functions}

The intensity function is $\Lambda(r, \theta) = \rho r^\epsilon$ for $\epsilon > -2$ which yields the following:
\begin{align}
n &= \frac{2\pi \rho}{2+\epsilon}R^{2+\epsilon}, \;\;g(n) = R = \left(\frac{2+\epsilon}{2\pi \rho}Nc\right)^{\frac{1}{2+\epsilon}}\\
H(x) &= 1-\Ind{1<x}\lim_{N\to\infty}\frac{ \left(\frac{2+\epsilon}{2\pi \rho}Nc\right)^{\frac{\epsilon}{2+\epsilon}} x^{-\frac{2+\epsilon}{\alpha}} }{ \left(\frac{2+\epsilon}{2\pi \rho}Nc\right)^{\frac{\epsilon}{2+\epsilon}}\,}\\
&=1-\Ind{1<x}x^{-\frac{2+\epsilon}{\alpha}}\\
\frac{dH(x)}{dx} &= \frac{2+\epsilon}{\alpha}x^{-\frac{2+\epsilon+\alpha}{\alpha}}\Ind{1<x}
\end{align}
which when substituted into \eqref{Eqn:FixedPoint} yields:
\begin{align}
&-\beta\sigma^2 + 1 = \beta c \int_{0}^\infty\frac{2+\epsilon}{\alpha(1+\beta x) }x^{-\frac{2+\epsilon}{\alpha}}\Ind{1<x} \, dx \\
& = \frac{(2+\epsilon)\beta c}{\alpha}\left[ \int_{0}^\infty\frac{x^{-\frac{2+\epsilon}{\alpha}}}{(1+\beta x) } \, dx -\int_{0}^1\frac{x^{-\frac{2+\epsilon}{\alpha}}}{(1+\beta x) } \, dx \right] \nonumber \\
& = \frac{(2+\epsilon) c}{\alpha}\beta^{\frac{2+\epsilon}{\alpha}}\pi\csc\left(\frac{2\pi+\pi\epsilon}{\alpha}\right)\nonumber \\
&-\frac{(2+\epsilon)\beta c}{\alpha-2-\epsilon}\,_2F_1\left(1, \frac{-2-\epsilon+\alpha}{\alpha}, \frac{-2-\epsilon+2\alpha}{\alpha}, -\beta x\right) \nonumber
\end{align}
which concludes the proof of Theorem \ref{Theorem:PowerLaw}.

\bibliographystyle{IEEEbib}

\bibliography{main}

\end{document}